\documentclass[aps,prl,superscriptaddress,amsmath,amssymb,twocolumn,amsfonts,floatfix, longbibliography,10pt]{revtex4-2}

\usepackage{graphicx}% Include figure files
\usepackage{dcolumn}% Align table columns on decimal point
\usepackage{bm}% bold math
\usepackage{array}
\usepackage[dvipsnames]{xcolor}
\definecolor{nblue}{rgb}{0.0, 0.0, 1.0}
\definecolor{magenta}{rgb}{0.79, 0.08, 0.48}
\definecolor{dgreen}{rgb}{0.0, 0.5, 0.0}
\usepackage[colorlinks,linkcolor=nblue,urlcolor=magenta,citecolor=dgreen,
plainpages=false,pdfpagelabels,breaklinks]{hyperref}
\usepackage{booktabs}
\usepackage[normalem]{ulem}
\usepackage{amsmath}
\usepackage{orcidlink}
\usepackage{dcolumn,color,booktabs,microtype,afterpage,chemformula}
\usepackage{soul}

\setstcolor{red}

\newcommand{\beq}{\begin{equation}}
\newcommand{\eeq}{\end{equation}}
\newcommand{\bea}{\begin{eqnarray}}
\newcommand{\eea}{\end{eqnarray}}

\newcommand{\eqn}[1] {Eq.~(\ref{#1})}
\newcommand{\fig}[1]{Fig.~\ref{#1}}

\begin{document}

\title{\textrm{Observation of Time-Reversal Symmetry Breaking in the Type-I Superconductor \ch{YbSb2}}}

\author{Anshu Kataria}\thanks{These authors contributed equally to this work}
\affiliation{Department of Physics, Indian Institute of Science Education and Research Bhopal, Bhopal, 462066, India}

\author{{Shashank Srivastava}\,\orcidlink{0009-0009-5065-4516}}\thanks{These authors contributed equally to this work}
\affiliation{Department of Physics, Indian Institute of Science Education and Research Bhopal, Bhopal, 462066, India}

\author{{Dibyendu Samanta}\,\orcidlink{0009-0004-3022-7633}}\thanks{These authors contributed equally to this work}
\affiliation{Department of Physics, Indian Institute of Technology, Kanpur 208016, India}

\author{{Pushpendra Yadav}\,\orcidlink{0000-0003-4784-9573}}\thanks{These authors contributed equally to this work}
\affiliation{Department of Physics, Indian Institute of Technology, Kanpur 208016, India}

\author{Poulami Manna}
\affiliation{Department of Physics, Indian Institute of Science Education and Research Bhopal, Bhopal, 462066, India}

\author{Suhani Sharma}
\affiliation{Department of Physics, Indian Institute of Science Education and Research Bhopal, Bhopal, 462066, India}

\author{Priya Mishra}
\affiliation{Department of Physics, Indian Institute of Science Education and Research Bhopal, Bhopal, 462066, India}

\author{Joel Barker}
\affiliation{Physics Department, University of Warwick, Coventry CV4 7AL, United Kingdom}

\author{Adrian D. Hillier}
\affiliation{ISIS Facility, STFC Rutherford Appleton Laboratory, Didcot OX11 0QX, United Kingdom}

\author{Amit Agarwal}
\affiliation{Department of Physics, Indian Institute of Technology, Kanpur 208016, India}

\author{{Sudeep Kumar Ghosh}\,\orcidlink{0000-0002-3646-0629}}
\email{skghosh@iitk.ac.in}
\affiliation{Department of Physics, Indian Institute of Technology, Kanpur 208016, India}

\author{Ravi Prakash Singh}
\email{rpsingh@iiserb.ac.in}
\affiliation{Department of Physics, Indian Institute of Science Education and Research Bhopal, Bhopal, 462066, India}

\date{\today}

\begin{abstract}
The spontaneous breaking of time-reversal symmetry is a hallmark of unconventional superconductivity, typically observed in type-II superconductors. Here, we report evidence of time-reversal symmetry breaking in the type-I superconductor YbSb$_2$. Zero-field $\mu$SR measurements reveal spontaneous internal magnetic fields emerging just below the superconducting transition, while transverse-field $\mu$SR confirms a fully gapped type-I superconducting state. Our first-principles calculations identify YbSb$_2$ as a ${\mathbb Z}_2$ topological metal hosting a Dirac nodal line near the Fermi level. Symmetry analysis within the Ginzburg–Landau framework indicates an internally antisymmetric nonunitary triplet (INT) state as the most probable superconducting ground state. Calculations based on an effective low-energy model further demonstrate that this INT state hosts gapless Majorana surface modes, establishing YbSb$_2$ as a topological superconductor. Our results highlight YbSb$_2$ as a unique material platform where type-I superconductivity coexists with triplet-pairing and nontrivial topology.
\end{abstract}

\maketitle

Topological quantum materials have emerged as a central theme in condensed matter physics due to their ability to host exotic boundary states exhibiting emergent quantum phenomena and offering potential applications in quantum technologies~\cite{Tkachov2022}. In particular, topological superconductors are predicted to host gapless Majorana excitations that can be utilized for fault-tolerant quantum computation~\cite{sato2017,Qi2011}. Realizing such unconventional states often requires the discovery of superconductors with nontrivial topological band structures or symmetry-protected degeneracies. Recently, superconductors possessing nonsymmorphic crystal symmetries have attracted growing attention as a promising platform for realizing topological superconductivity, as these symmetries can enforce nontrivial normal state band topology coexisting with superconductivity~\cite{Huang2018,Kim2019,Yadav2024,Gao2020}.

Despite extensive investigations, only a limited number of nonsymmorphic superconductors are known, such as InBi$_2$~\cite{Kuang2021}, Tl$_{2-x}$Mo$_6$Se$_6$~\cite{Huang2018}, $A15$ compounds Ta$_3\textit{X}$ ($\textit{X}$=Sb, Sn, Pb)~\cite{Kim2019}, Ti$_3\textit{X}$ ($\textit{X}$=Ir, Sb)~\cite{Ti3X}, $\textit{X}$RuB$_2$ ($\textit{X}$=Y, Lu)~\cite{Gao2020,YRuB2,RRuB2}, $\textit{M}$IrGe ($\textit{M}$=Ti, Hf, Zr)~\cite{Meena2025nonsym,Meena2025superconductivity} and Zr$_2$Ir \cite{Zr2Ir}. Particularly intriguing is the subset of these materials that breaks time-reversal symmetry (TRS) in the superconducting state, a group that includes LaNiGa$_2$~\cite{Hillier2012,Ghosh2020b,Badger2022}, La(Pt,Ni)(Si,Ge)~\cite{Shang2022Weyl,LaPtGe,LaMSi}, $\textit{X}$OsSi ($\textit{X}$ = Ta, Nb)~\cite{Ghosh2022Dirac}, ${\mathrm{Zr}}_{3}\mathrm{Ir}$~\cite{Shang2020,Zr3Ir}, HfRhGe~\cite{Sajilesh2025time}, LaPt$_3$P~\cite{Biswas2021chiral} and CaSb$_2$~\cite{casb2trsb,casb2arpes}. The observation of TRS-breaking in superconductors is usually confined to systems with complex crystal structure, including cage-type and noncentrosymmetric, or the presence of strong correlations~\cite{Ghosh2020a,Shan2022,Ghosh2021time,bhattacharyya2018unconventional,bhattacharyya2015unconventional,bhattacharyya2015broken}, and all known systems to date exhibit type-II superconducting character. Meanwhile, investigations of type-I superconductors with centrosymmetric and noncentrosymmetric structures, such as AuBe~\cite{Singh2019,Beare2019}, $A$IrGe ($A$ = Hf, Zr, Ti)~\cite{Meena2025superconductivity,Meena2025nonsym}, La$M$Si$_3$ ($M$ = Ir, Rh, Pd)~\cite{Anand2011,Anand2014,Smidman2014}, Pb$_2$Pd~\cite{Arushi2021}, and SnAs \cite{SnAs} have consistently revealed conventional, TRS-preserving singlet pairing. Furthermore, the observed TRS-breaking in elemental Re~\cite{Shang2018} is still under discussion, due to the possible effect of muon diffusion in the relaxation process, as suggested in a recent investigation~\cite{Jonas2022}. These observations point to a fundamental open question: Can TRS be spontaneously broken in a type-I superconducting state, and under what electronic or symmetry conditions might such a phase emerge?

{Here}, we address this question {by demonstrating a TRS broken type-I superconducting state} in YbSb$_2$, a nonsymmorphic square-net compound belonging to the $\textit{R}$Sb$_2$ ($\textit{R}$ = Ca, Yb) family~\cite{Ikeda2020,Oudah2022,Bodnar1967}. We suggest that YbSb$_2$ is a topological metal ${\mathbb{Z}}_2$ \cite{Ortiz2020,Ghosh2023zrossi} in the normal metallic state, providing a clean material platform for exploring topological superconductivity. From zero and transverse-field muon spin relaxation and rotation ($\mu$SR) measurements, we observe a clear, reproducible time-reversal symmetry–breaking signal with a type-I superconducting state. This establishes YbSb$_2$ as the first known type-I superconductor with broken TRS. The fully gapped superconducting state is further supported by the specific heat measurements. Guided by symmetry analysis, we propose that the most plausible superconducting ground state in YbSb$_2$ is an internally antisymmetric nonunitary triplet (INT) state with nontrivial topology characterized by Majorana surface modes.

\begin{figure*}[!t]
\centering 
\includegraphics[width=0.985\textwidth, origin=b]{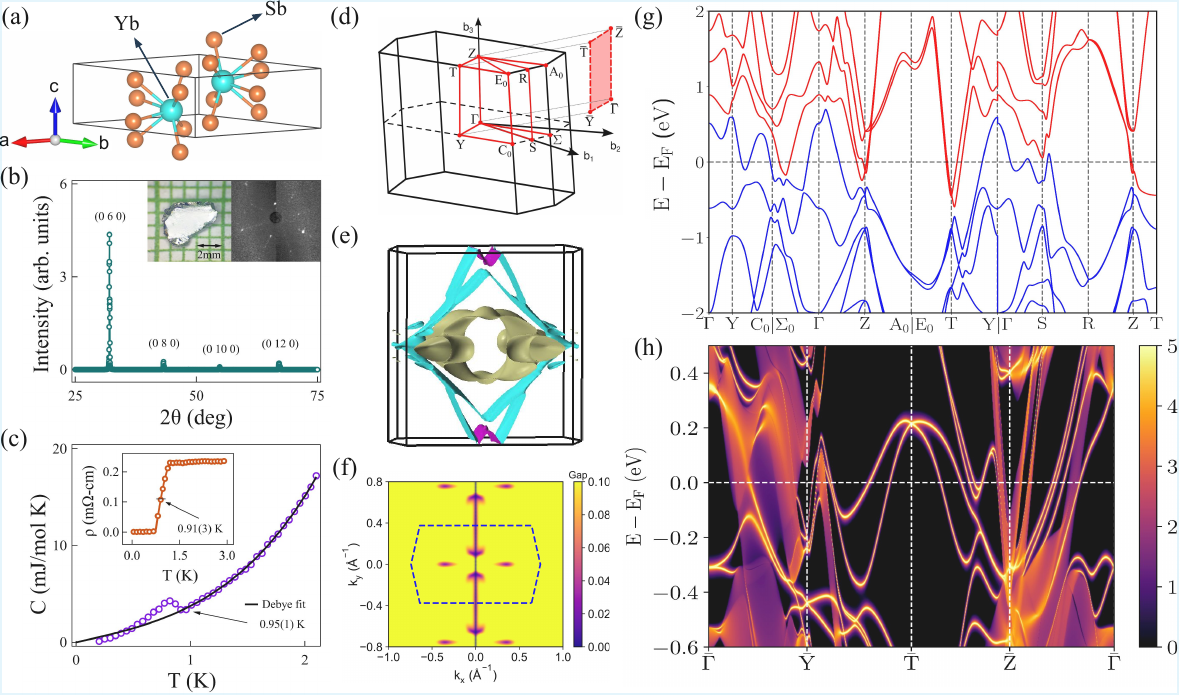}
\caption{\label{fig1} \textbf{Sample characterization and electronic band structure of YbSb$_2$:} (a) Schematic of a unit cell of YbSb$_2$.
(b) XRD spectrum for the single crystals. The inset shows the as-grown single crystal (left) and the Laue pattern (right). 
(c) Temperature-dependent specific heat data indicating $T_c$ and the normal state data fitting. Inset: Zero drop in the resistivity data showing $T_c$ at the midpoint.
(d) The first Brillouin zone, including the $(1\bar{1}0)$ surface-Brillouin zone and the high symmetry directions. 
(e) Fermi surfaces with SOC composed of one conduction, and two valence bands. 
(f) Nodal line along the Z-T path of the bulk BZ representing four-fold degeneracy due to doubly-degenerate top two valence bands VB1 and VB2 near the Fermi level.
(g) Electronic band structure with SOC, where SOC introduces an overall direct band gap between the conduction (red) and valence (blue) bands.
(h) Bulk and surface spectral functions calculated for the $(1\bar{1}0)$ surface.}
\end{figure*}

\noindent \textit{Structural characterization and bulk superconductivity:} 
Single-crystalline YbSb$_2$ samples were synthesized by the modified Bridgman method. Room-temperature X-ray diffraction (XRD) confirms that YbSb$_2$ crystallizes in the nonsymmorphic space group $Cmcm$ [Fig.~\ref{fig1}(b)]. The crystal structure contains two formula units per unit cell, forming distinct quintuple layers within the $ac$ plane, as shown in Fig.~\ref{fig1}(a). These quintuple layers are stacked along the crystallographic $b$ axis and are held together by weak van der Waals interactions. The Laue diffraction and XRD pattern of the single crystals confirm the high-quality crystallinity of the sample [see the inset of Fig. \ref{fig1}(b)].

YbSb$_2$ is a {known} type-I bulk superconductor, {characterized via} resistivity, specific heat and magnetic susceptibility measurements \cite{Zhao2012,Gao2025}. {We} verify the superconducting properties of our sample using complementary electrical resistivity and heat capacity measurements. The zero drop in resistivity and the superconducting jump in specific heat, presented in Fig. \ref{fig1}(c), confirm the bulk superconductivity with a superconducting transition at the critical temperature $T_c=0.95(1)$ K. Our specific heat data show a single superconducting transition, unlike the possibility of a second transition in the previous report \cite{Zhao2012}. The electronic specific data in the superconducting state, fitted with the s-wave model, reveal an isotropic superconducting gap with weak electron-phonon coupling [see the Supplemental Material (SM) \cite{SM}]. The type-I nature of the superconducting ground state in YbSb$_2$ is further corroborated using the $\mu$SR measurements.

\begin{figure*}[!t]%{r}{0.5\textwidth}
\centering 
\includegraphics[width=0.985\textwidth, origin=b]{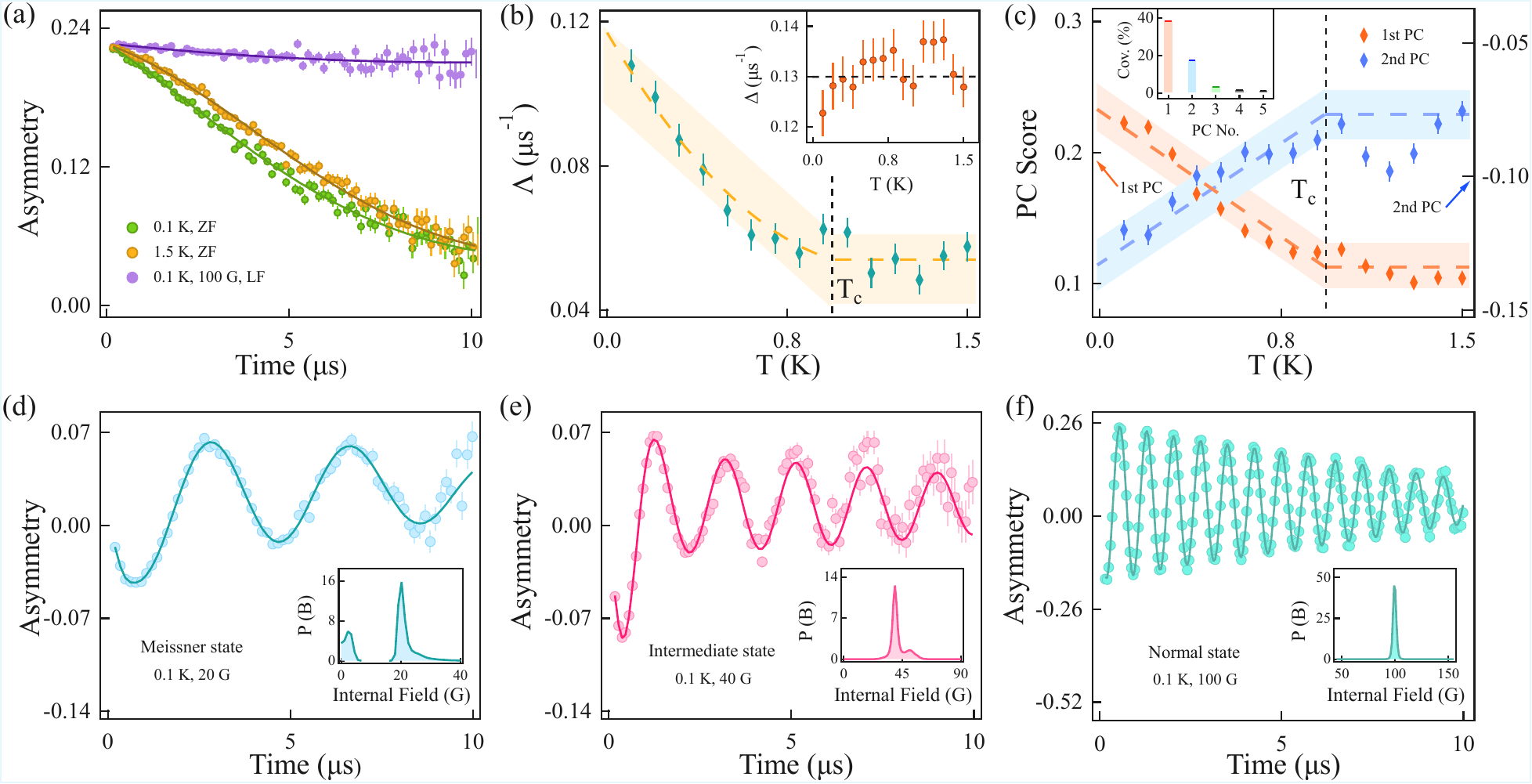}
\caption{\label{Fig:ZF} \textbf{Time reversal symmetry breaking and type-I superconductivity in YbSb$_2$:}
(a) Zero-field asymmetry spectra of YbSb$_2$, above (1.5 K) and below (0.1 K) the $T_c$, with longitudinal-field spectra at 100 G, 0.1 K, where the solid line indicates the corresponding fits. (b) Electronic relaxation rate ($\Lambda$) versus temperature indicates a significant increase below the superconducting transition temperature. Inset: The nuclear relaxation rate ($\Delta$) {is almost constant with} temperature. (c) The variation of 1st and 2nd {principal component} scores with temperature. Inset shows covariance percentage captured by different principal components. The shaded bands and dashed lines are guides to the eye. Transverse-field asymmetry spectra at 0.1 K for different applied magnetic fields (d) 20 G, (e) 40 G, and (f) 100 G. The probability distribution of the magnetic field for the respective spectra is shown in the corresponding inset, indicating different states.}
\end{figure*}

\noindent \textit{Electronic band structure and topological properties:} To understand the normal state characteristics of YbSb$_2$, we performed first principles calculations within density functional theory (DFT). {We used the generalized gradient approximation (GGA)~\cite{Perdew1996} as implemented in the Vienna \textit{ab initio} simulation package (VASP)~\cite{vasp-1,vasp-2}} [details are in the SM \cite{SM}]. The bulk and $(010)$-surface Brillouin zones (BZs) of YbSb$_2$ are shown in Fig.~\ref{fig1}(d). We treat the electronic states Yb-4f within the core, as these are located approximately 0.7~eV below the Fermi level and are fully occupied~\cite{Sato1999}. {Figure} \ref{fig1}(g) shows the electronic band structure of YbSb$_2$ with spin-orbit coupling (SOC). Without SOC, three doubly degenerate bands cross the Fermi energy ($E_F$), confirming its metallic nature [see SM \cite{SM}]. SOC induces significant band splitting (up to $\sim 80$ meV), creating a continuous direct band gap between the valence (blue) and conduction (red) {bands} near $E_F$. However, the material remains metallic due to highly dispersive low-energy bands forming sizable electron and hole pockets. Kramers theorem ensures that all bands remain at least twofold degenerate in this centrosymmetric system. YbSb$_2$ is inherently multi-band, with Sb-$p$ and Yb-$d$ orbitals dominating the density of states at $E_F$. {Figure} \ref{fig1}(e) shows the Fermi surfaces with SOC, revealing a quasi-2D character with nearly parallel sheets across extended regions of the BZ, consistent with the layered structure of  YbSb$_2$~\cite{Sato1999}. Interestingly, we find a fourfold degenerate nodal line, protected by nonsymmorphic glide mirror symmetry, emerges between the first and second valence band pairs and the first and second conduction band pairs along the high-symmetry $Z-T$ direction as shown in \fig{fig1}(f).

The continuous direct gap between the conduction and valence bands across the entire BZ in YbSb$_2$ allows for a topological classification similar to a gapped insulator~\cite{Schoop2015,Nayak2017,Ortiz2020}. Our calculations reveal a band inversion between the orbitals Yb-\(d\) and Sb-\(p\) at the $\Gamma$ point of the BZ with SOC, suggesting a possible topologically nontrivial character~\cite{Bansil2016} [see SM \cite{SM}]. Using the centrosymmetry of YbSb\(_2\), we compute topological invariants ${\mathbb{Z}}_{2}$ using two independent methods: (i) Wannier charge center (WCC) approach and (ii) Product of parity eigenvalues, considering the filled valence bands up to the highest green band in \fig{fig1}(g). Both methods yield a topological invariant of $(1; 1 1 0)$, confirming that YbSb\(_2\) is a ${\mathbb{Z}}_{2}$ topological metal, which is {topologically equivalent to a doped strong ${\mathbb{Z}}_{2}$ topological insulator (TI)}~\cite{Schoop2015,Nayak2017,Ortiz2020}.

To further characterize the topology of YbSb$_2$, we compute the spectral function of a finite slab cleaved along the layered plane ($1\bar{1}0$) of the primitive unit cell~\cite{Wu2018}. \fig{fig1}(h) shows spectral contributions from both the bulk and the surface states along high-symmetry paths of the ($1\bar{1}0$) surface Brillouin zone. Several intense surface states appear within the bulk energy gap, including a nontrivial surface Dirac cone at $\bar{Y}$ ($\sim -0.48$ eV) and another at $\bar{T}$ ($\sim 0.25$ eV), consistent with the characteristics of a $\mathbb{Z}_2$ topological metal~\cite{Schoop2015,Nayak2017,Ortiz2020}. Additionally, nontrivial surface states connecting the valence and conduction bands disperse across the Fermi level, making them experimentally accessible via techniques such as angle-resolved photoemission spectroscopy (ARPES).

\noindent \textit{TRS-breaking superconductivity from zero-field $\mu$SR}: The time-dependent zero-field (ZF)-$\mu$SR asymmetry spectra in the normal and superconducting states of YbSb$_2$ are presented in \fig{Fig:ZF}(a). In the superconducting state ($T=0.1$~K), the spectra exhibit a noticeable increase in the relaxation rate compared to the normal state ($T=1.5$ K), indicating the emergence of spontaneous magnetic fields {in} the superconducting state. Moreover, from the longitudinal-field (LF)-$\mu$SR spectra in \fig{Fig:ZF}(a), we note that applying a small longitudinal magnetic field of $\sim$10~mT is sufficient to decouple the muon spins from the relaxation channel, giving rise to a static muon depolarization spectrum.

To quantitatively analyze the asymmetry spectra, we fit the ZF-$\mu$SR asymmetry spectra using a damped Gaussian Kubo-Toyabe form with an additional exponential relaxation term and a flat background contribution~\cite{kt}:
\beq
A(t)= A_0 \left[\frac{1}{3} + \frac{2}{3}(1-\Delta^2t^2)e^{-\frac{\Delta^2 t^2}{2}}\right]e^{-\Lambda t} + A_{bg}.
\label{eqn:ZF}
\eeq
Here, $A_0$ and $A_{bg}$ represent the initial and background asymmetries, respectively. The parameter $\Delta$ corresponds to the relaxation rate associated with static nuclear dipolar fields, while $\Lambda$ accounts for the additional electronic relaxation rate. The temperature variation of the fitted parameters $\Lambda$ and $\Delta$ is shown in \fig{Fig:ZF}(b). A systematic increase in $\Lambda$ with decreasing temperature suggests the emergence of a spontaneous magnetic field below $T_c$, whereas $\Delta$ is temperature independent.

Thus, the combined ZF- and LF-$\mu$SR results demonstrate the emergence of a static or quasistatic spontaneous magnetic field in the superconducting state of YbSb$_2$, establishing spontaneous TRS-breaking at $T \lesssim T_c$. The magnetic field in the superconducting state of YbSb$_2$ estimated using the change in $\Lambda$ between $T=0$ K and $T>T_c$ 
from the relation: $d\Lambda = \gamma_{\mu} B_{in} $ is $B_{in} \approx 0.44(3)$~G, where $\gamma_{\mu}/2\pi$ = 13.55 kHz/G is the muon gyromagnetic ratio. The magnitude of the magnetic field is of similar order as for other TRS-breaking superconductors~\cite{Ghosh2020a}. 

To support our ZF-$\mu$SR analysis of YbSb$_2$, we applied the principal component analysis (PCA) [details in SM \cite{SM}]. A joint PCA with established TRS-breaking materials shows that the first two PCs dominate the variance [see the scree plot in the inset of \fig{Fig:ZF}(c)], with PC1 contributing most strongly~\cite{tula2022joint}. The error bars on the PC scores were obtained by propagating uncertainties in the asymmetry curves, and their temperature evolution [\fig{Fig:ZF}(c)] reveals marked changes in PC1 and PC2 for $T \lesssim T_c$. These trends are consistent with our analysis based on the Kubo–Toyabe formula [\eqn{eqn:ZF}] and confirm the appearance of spontaneous internal magnetic fields below $T_c$.

\noindent \textit{Type-I superconductivity from transverse-field $\mu$SR}:
The transverse-field (TF)-$\mu$SR asymmetry spectra of YbSb$_2$ under different applied magnetic fields are shown in \fig{Fig:ZF}(d)-(f), with the corresponding insets displaying the probability distribution of the internal field, $P(B)$. In an applied field of $20$~G and $T =0.1$~K [\fig{Fig:ZF}(d), inset], $P(B)$ exhibits a peak at $B \sim 4$~G, corresponding to nuclear dipolar fields in the Meissner state, while a second peak at $B \sim20$~G originates from muons stopping in the sample holder. In the applied field 40~G, the field distribution [\fig{Fig:ZF}(e), inset] reveals a secondary peak at $B \sim51$~G, indicating the coexistence of superconducting and normal domains in an intermediate state. The fitting of the TF-$\mu$SR spectra includes the background contribution of the silver holder [see SM \cite{SM}]. The presence of an internal field higher than the applied field suggests a critical field of 51~G due to demagnetization effects, a characteristic feature of type-I superconductors. This contrasts with type-II superconductors, where a lower-field peak would typically indicate a flux-line lattice. Additionally, the TF-$\mu$SR spectra show a reduction in initial asymmetry, similar to other type-I superconductors, which arises due to detector limitations in detecting positrons emitted along the muon spin polarization in the TF-geometry \cite{Anand2011,lns,Singh2019,Beare2019}. In the normal state in an applied field $B \sim 100$~G [\fig{Fig:ZF}(f)], the field distribution shows a single sharp peak. This indicates complete penetration of the field into the bulk and confirms the destruction of the superconducting state.

Our thermodynamic and $\mu$SR measurements do not indicate the presence of any secondary phase in YbSb$_2$. TF-$\mu$SR measurements do not show any crossover between type-I and type-II superconducting phases or a mixed type-I / II (type-1.5) superconducting phase \cite{t1.5}. Furthermore, considering the presence of possible nontrivial topological states, the observation of secondary higher upper critical field via the Radio-frequency tunnel diode oscillator (TDO) might be due to the surface superconductivity. This effect can be ascribed to the surface states of YbSb$_2$, which have not been previously considered \cite{Zhao2012}. As noted, TDO is a surface-sensitive technique, whereas specific heat and $\mu$SR are bulk probes.

\noindent \textit{Topological superconductivity in YbSb$_2$:} To understand the nature of the superconducting ground state in YbSb$_2$, we employ a Ginzburg–Landau theory based symmetry analysis~\cite{Ghosh2020a,Annett1990,sigrist1991}. The low-symmetry orthorhombic point group D$_{2h}$ of YbSb$_2$ admits only 1D irreducible representations. This excludes any symmetry-allowed TRS-breaking superconducting instability in the strong SOC limit. In the weak SOC regime, the D$_{2h}\otimes$SO(3) symmetry of the normal state can, in principle, host TRS-breaking superconducting instabilities; however, all such states possess point nodes~\cite{Hillier2012,Weng2016}, inconsistent with the experimentally observed fully gapped superconductivity in YbSb$_2$. Our first-principles calculations indicate strong SOC effects, which further rule out these nodal TRS-breaking possibilities. Although nonsymmorphic symmetries can enforce nodal superconductivity along high-symmetry directions at the Brillouin-zone boundary, they cannot generate a multicomponent order parameter required for TRS-breaking~\cite{sumita2018}. Therefore, all possible symmetry-allowed single-band superconducting instabilities in YbSb$_2$ are incompatible with experimental observations.

Our first-principles results also reveal that multiple orbitals contribute significantly to the density of states at the Fermi level in YbSb$_2$. This motivates us to consider an internally antisymmetric nonunitary triplet (INT) superconducting state, previously proposed for LaNiGa$_2$, which possesses the same point-group symmetry~\cite{Hillier2012,Weng2016}. In this state, pairing occurs between electrons on the same site but belonging to different orbitals, with antisymmetry enforced in orbital space. The corresponding pairing potential is $\hat{\Delta}=(i\tau_y)\otimes(\mathbf{d}\cdot\mathbf{s})(i\sigma_y)$, where the triplet $\mathbf{d}$-vector is $\mathbf{d}=\Delta_0\boldsymbol{\eta}$, with $|\boldsymbol{\eta}|^2=1$ and $\Delta_0$ is the uniform pairing amplitude \cite{Hillier2012,Weng2016}. The nonunitarity of this state is characterized by $\mathbf{q}=i(\boldsymbol{\eta}\times\boldsymbol{\eta}^*)\neq0$. The presence of parallel and nearly degenerate Fermi surface sheets in YbSb$_2$ provides favorable conditions to stabilize the INT state.

\begin{figure}[!t]
\centering
\includegraphics[width=0.95\columnwidth]{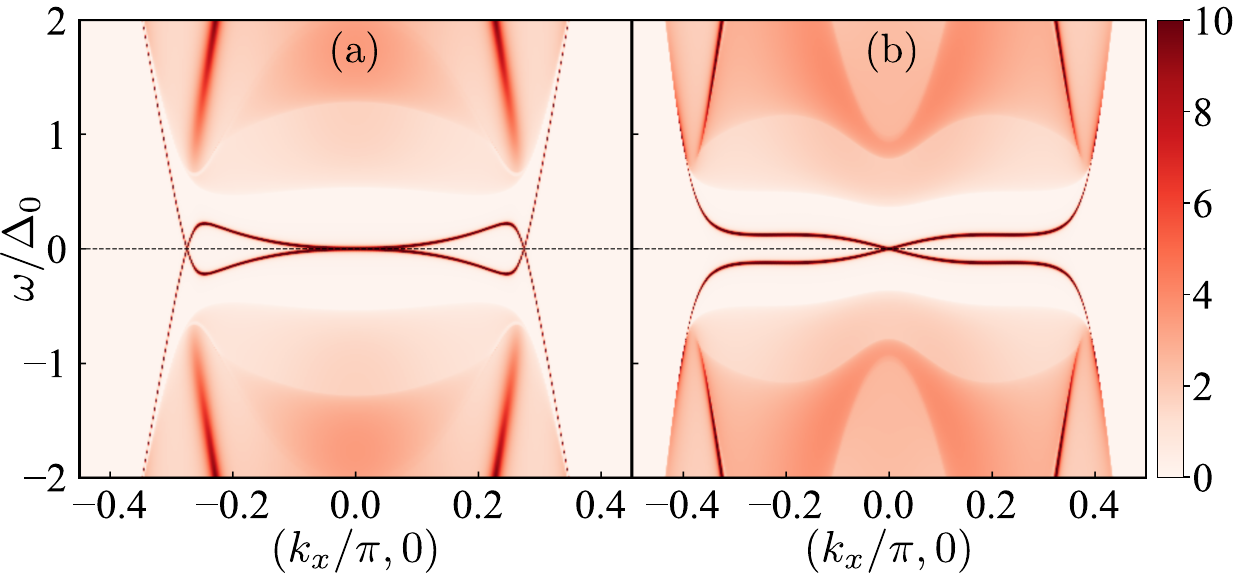}
\caption{\textbf{Nontrivial topology of the internally antisymmetric nonunitary triplet superconducting state:} Surface spectral function for the internally antisymmetric nonunitary triplet state of a $\mathbb{Z}_2$ topological metal for two different values of the chemical potential (a) $\mu=0.90$ and (b) $\mu=1.2$, with $\Delta_0=0.1$ and $\mathbf{\eta}=(1/\sqrt{2})(1, e^{i\pi/4},0)$. The parameters chosen in the Hamiltonian in \eqn{eqn:dirac_eq} are $b=0.5$, $v=1.0$ and $m=-0.7$. The surface states display a pronounced twisting dispersion, featuring an additional crossing away from the $\Gamma$-point that becomes more prominent with increasing $\mu$.}
\label{spectral_function}
\end{figure}

Our first-principles calculations reveal that $\mathrm{YbSb}_2$ is a $\mathbb{Z}_2$ topological metal, equivalent to a doped strong $\mathbb{Z}_2$ TI. Consequently, the low-energy normal-state physics of $\mathrm{YbSb_2}$ can be captured by a general three-dimensional massive Dirac Hamiltonian~\cite{Hasan2010,Qi2011,Chiu2016,Bansil2016,Altland1997,Kane2005,Bernevig2006,Fu2007,Liu2010TI,Ryu_2010}:
\begin{equation}
    \hat{h}_{\mathbf{k}}=\epsilon_0 - \mu + \sum_{i=1}^{3} \left(a_i k_{i}^2  + v_i k_i \Gamma_{1i} \right) + \left( m + \sum_{i=1}^{3} b_i k_{i}^2 \right) \Gamma_{30}. \label{eqn:dirac_eq}
\end{equation}
\noindent Here, $\Gamma_{\alpha\beta} = \tau_\alpha \otimes \sigma_\beta$, where $\tau_{\{1,2,3\}}$ and $\sigma_{\{1,2,3\}}$ are Pauli matrices acting on orbital and spin degrees of freedom respectively, while $\tau_0$ and $\sigma_0$ represent the corresponding identity matrices. $v_i$ denote the Fermi velocities along the spatial directions $i$, $\mu$ is the chemical potential, $m$ represents the isotropic Wilson–Dirac mass term, and $\epsilon_0$ introduces the particle-hole asymmetry. Assuming cubic symmetry and for simplicity, we take $v_i = v$, $b_i = b$ and $a_i = \epsilon_0 = 0$. This Hamiltonian respects parity, time-reversal , and charge-conjugation symmetries. For $m b > 0$, the system lies in a topologically trivial phase with a minimum spectral gap of $2m$, while for $m b < 0$, it enters a topologically nontrivial phase.

The quasiparticle density of states, computed using the Bogoliubov-de Gennes formalism for the INT superconducting state built on the normal-state Hamiltonian in Eq.~\eqref{eqn:dirac_eq}, exhibits a full gap consistent with experimental observations in $\mathrm{YbSb}_2$ [see SM \cite{SM}]. To investigate the topological nature of the INT superconducting state, we further evaluate the surface spectral function $A(\mathbf{k},\omega)=-\frac{1}{\pi}{\rm Im}\left[ {\rm Tr}G(\mathbf{k},\omega) \right]$ using the transfer-matrix formalism [see SM \cite{SM}], where $G(\mathbf{k},\omega)$ is the retarded surface Green’s function. The behavior of the spectral function for two representative values of the chemical potential ($\mu$) is shown in Fig.~\ref{spectral_function}(a,b). The emergence of zero-energy surface Andreev bound states (SABS) demonstrates that the INT superconducting state in $\mathrm{YbSb}_2$ is topologically nontrivial~\cite{Hao_2011,Hsieh_2012}. A linearly dispersed Majorana mode appears at the $\Gamma$ point, characteristic of a three-dimensional topological superconductor. Moreover, the SABS exhibit a distinct twisting dispersion in the surface Brillouin zone, connecting the Majorana branch at $\Gamma$ to a secondary crossing away from $\Gamma$, similar to Cu-doped Bi$_2$Se$_3$ topological superconductor~\cite{Hsieh_2012}. As the chemical potential increases, this secondary crossing gradually opens up, signaling a topological phase transition driven by tuning of the chemical potential~\cite{Hao_2011,Hsieh_2012}.

\noindent\textit{Summary and Conclusions:} 
Our study establishes YbSb$_2$ as a {single phase} type-I bulk superconductor{. It has} a critical temperature $T_c = 0.95(1)$ K and a critical field $H_c = 51(1)$ G at 0.1 K. Zero-field $\mu$SR measurements reveal a clear TRS-breaking signal below $T_c$, identifying YbSb$_2$ as the first type-I superconductor known to exhibit spontaneous TRS-breaking. Transverse-field $\mu$SR confirms a fully gapped superconducting state, which is supported by specific heat measurements. Electronic structure calculations show that YbSb$_2$ is a $\mathbb{Z}_2$ topological metal. The multiband character of YbSb$_2$ necessitates a symmetry analysis beyond a single-band description; within the Ginzburg–Landau framework, the internally antisymmetric nonunitary triplet (INT) state emerges as the most plausible superconducting ground state. This state is topologically nontrivial and {supports} Majorana surface modes, expected to {manifest} as zero-bias conductance peaks in tunneling or point-contact spectroscopy \cite{Hao_2011,Hsieh_2012}. These results establish YbSb$_2$ as a rare platform where type-I superconductivity, triplet pairing, and topology coexist. Future complementary investigations utilizing $\mu$SR, ARPES, and STM, combined with a systematic exploration of the $R$Sb$_2$ ($\textit{R}$ = Ca, Yb) family and related nonsymmorphic compounds, will be essential to elucidate the microscopic origin of TRS breaking and its connection to topological superconductivity.

\noindent\textit{Acknowledgments:} A.K. acknowledges the funding agency Council of Scientific and Industrial Research (CSIR), Government of India, for providing a SRF fellowship [Award No. 09/1020(0172)/2019-EMR-I]. R.~P.~S. acknowledges the SERB, Government of India, for the Core Research Grant No. CRG/2023/000817. S.~K.~G. also acknowledges financial support from Anusandhan National Research Foundation (ANRF), erstwhile SERB, Government of India, via the Startup Research Grant: SRG/2023/000934. The authors thank ISIS, STFC, UK for providing beamtime to conduct the $\mu$SR experiments. We acknowledge the high-performance computing facility at IIT Kanpur and the National Supercomputing Mission (NSM) for providing computing resources of `PARAM Sanganak' at IIT Kanpur, which is implemented by C-DAC and supported by the Ministry of Electronics and Information Technology (MeitY) and DST, Government of India. Sh. S. and P. Y. acknowledge the UGC, India, for a Senior Research Fellowship. We thank Barun Ghosh and Atasi Chakraborty for many fruitful discussions.

\bibliography{Library}

\end{document}